# Physics-informed extreme learning machine for Terzaghi's consolidation problems and interpretation of coefficient of consolidation based on CPTu data

Dr He **Yang**

E-mail: yanghesdu@mail.sdu.edu.cn

School of Qilu Transportation, Shandong University, Jinan, 250002, China

Prof Pin-Qiang **Mo**

Email: pinqiang.mo@cumt.edu.cn

School of Mechanics and Civil Engineering, China University of Mining and Technology, Xuzhou, 221116, China

Dr Fei **Ren**

Email: ren87@outlook.com

Faculty of Mechanical Engineering, Qilu University of Technology (Shandong Academy of Sciences), Jinan, 250353, China

Professor Xueyu **Geng**

Email: Xueyu.Geng@warwick.ac.uk

School of Engineering, University of Warwick, Coventry, CV4 7AL, UK

Prof Hai-Sui **Yu**

Email: H.Yu@leeds.ac.uk

School of Civil Engineering, University of Leeds, Leeds, LS2 9JT, UK

Professor Pei-Zhi **Zhuang**

Corresponding author

E-mail: zhuangpeizhi@sdu.edu.cn

School of Qilu Transportation, Shandong University, Jinan, 250002, China

## Abstract

This paper conducts a preliminary study to investigate the feasibility of a physics-informed extreme learning machine (PIELM) for solving the Terzaghi consolidation equation and interpreting the coefficient of consolidation of soil from piezocone penetration tests (CPTu). In the PIELM framework, the target solution is approximated by a single-layer feed-forward extreme learning machine (ELM) network, instead of the deep neural networks typically employed in physics-informed neural networks (PINNs). Physical laws and measured data are integrated into a loss vector, which is minimized via least squares methods during ELM training. As a result, training efficiency is significantly improved by avoiding the gradient-descent optimisation commonly used in PINNs. The performance of PIELM is evaluated using three forward-problem case studies. Notably, a time-stepping strategy is incorporated into the PIELM framework to alleviate sharp gradients caused by inconsistent initial and boundary conditions. This paper further applies PIELM to estimate the soil consolidation coefficient, given that initial distributions of excess water pressure are often unavailable in CPTu dissipation tests (conducted following the pauses of penetration). By combining physical laws (excluding initial conditions) with measured data (i.e., excess pore-water pressure at the probe surface), the results demonstrate that PIELM is an effective tool for interpreting CPTu dissipation tests, owing to its ability to fuse data with physical constraints. This study contributes to the interpretation of consolidation coefficients from CPTu dissipation tests, particularly in scenarios where initial distributions of excess water pressure are not prior-known.

## 1. Introduction

Consolidation of saturated soils is a time-dependent process in which excess water pressure gradually dissipates and the corresponding effective stress increases. Understanding soil consolidation behaviour is essential for the design, construction and maintenance of geo-structures such as embankment settlement, slope stability and underground deformation, where settlement and stability are key concerns. Terzaghi (1943) was the first to mathematically describe this process by proposing the well-known one-dimensional consolidation theory. Biot (1941) further developed a three-dimensional consolidation theory that incorporates hydro-mechanical coupling to account for both pore water flow and soil deformation. Nowadays, more advanced consolidation theories have been proposed for the analysis of complex soil consolidation behaviour, which are able to take more factors into account, such as nonlinear compressibility, creep behaviour, and unsaturated soil states (Yin and Graham 1989; Yin and Graham 1996; Xie et al. 2002; Chen and Yin 2023).

The recent development of physics-informed machine learning (PIML) has made it possible to solve ordinary differential equations (ODEs) and partial differential equations (PDEs) by implementing physical laws (Raissi et al. 2019; Karniadakis et al. 2021). The most trending branch may be the physics-informed neural networks (PINNs), which enable various ODEs/PDEs to be solved with the universal function approximation capabilities of deep neural networks. Compared to traditional numerical methods, PINNs can more conveniently integrate physical laws with measured data, thereby giving PINNs a significant advantage in physics-data fusion and inverse analyses. Recent years have witnessed PINNs being widely used in geomechanics and geotechnical engineering. In particular, many studies applied PINNs to solve forward and inverse consolidation problems (Bekele 2021; Mandl et al. 2023; Sadiku 2023; Bekele 2024; Guo and Yin 2024; Lan et al. 2024a; Lan et al. 2024b; Wang et al. 2024; Xie et al. 2024; Yuan et al. 2024; Zhang et al. 2024a; Zhang et al. 2024b; Zhou et al. 2025). While these applications showed great success, one issue remains that the computational efficiency of PINNs is usually low (e.g., several minutes for one-dimensional cases). The multi-layered deep neural network and the time-consuming gradient descent methods for training network parameters could be responsible for the efficiency of conventional PINN. To improve training efficiency, an alternative is to replace fully connected neural networks with the extreme learning machine (ELM) (Huang et al. 2006), leading to the physics-informed extreme learning machine (PIELM) framework (Dwivedi and Srinivasan 2020). Compared to conventional PINNs, there is only one hidden layer in the PIELM framework, whose input layer parameters are fixed during training and the output layer parameters are directly calculated with the Moore–Penrose generalized inverse. As a result, both the training efficiency and accuracy can be significantly improved in solving PDEs (Calabrò et al. 2021; Dong and Li 2021; Liu et al. 2023; De Florio et al. 2024; Ren et al. 2025). For example, Ren et al. (2025) applied PIELM to solve Stefan problems, which saved more than 98% of training time and improves accuracy by a factor of $10^2$~$10^4$ than PINN.

The piezocone penetration test (CPTu) is one of the most widely used in-situ methods for determining the coefficient of consolidation of soils, owing to its convenience and low cost. During a CPTu test, a probe gradually penetrates the soil and excess water pressure is generated. This stage is referred to as the penetration stage. Then, a holding test is performed at a specified depth to record the dissipation of excess pore water pressure with time, and this phase is known as the dissipation stage. The dissipation rate of excess water pressure is closely related to soil permeability, and the dissipation stage is commonly formulated using the Terzaghi consolidation equation. It serves as the theoretical basis for determining the coefficient of consolidation from CPTu tests (i.e., the time-dependent dissipation of excess pore water pressure). Several empirical methods and trial-and-error curve-fitting approaches have been developed to interpret CPTu dissipation data (Soderberg 1962; Torstensson 1977; Baligh and Levadoux 1986; Gupta and Davidson 1986; Teh 1987). On the one hand, empirical methods are often region-specific and prone to human-induced uncertainties. On the other hand, trial-and-error curve-fitting methods are typically time-consuming and strongly dependent on the assumed initial distribution of excess pore water pressure at the beginning of the dissipation stage (i.e., the end of the penetration stage). Unfortunately, the initial distribution of excess pore water pressure cannot be directly measured by CPTu, if not practically impossible, and approximate expressions are commonly adopted with the aid of cavity expansion solutions. These expressions for initial excess pore water pressure suffer from several limitations: (i) they vary significantly among different soil types; and (ii) they are generally derived under the assumption of fully undrained conditions, thereby neglecting the dissipation of excess pore water pressure during the penetration stage. A question will arise: is it possible to infer the coefficient of consolidation from CPTu data without requiring the initial distribution of excess pore water pressure? From the perspective of machine learning, physics-informed machine learning (PIML) offers a potential framework to address this issue through the fusion of partial observational data and partial physical knowledge. Recently, Li et al. (2024) employed a physics-informed neural network (PINN) to estimate the in-situ coefficient of consolidation from CPTu tests. While they formulated the initial distribution of excess water pressure by a cavity expansion solution (Li et al. 2017), it remains unclear whether PIML can interpret CPTu data without prescribing initial conditions, and what the associated prediction accuracy would be.

To fill this gap, this paper presents a preliminary study aiming to: (i) explore the feasibility of PIELM in solving the Terzaghi consolidation equation; and (ii) investigate whether the coefficient of consolidation can be inferred from CPTu dissipation tests without initial distributions of excess pore water pressure. The rest part of this paper is organised as follows. First, Terzaghi consolidation equation as well as initial and boundary conditions is shown, and PINN framework for solving the consolidation equations is briefly summarised. Then the PIELM framework is detailed. Later, the performance of PIELM is validated using 3 case studies. Later, PIELM is applied to the interpretation of CPTu tests. Finally, main conclusions are drawn.

# 2. Preliminaries of Terzaghi's Consolidation Theory and PINN framework

Before showing the PIELM framework, the governing equations and the PINN framework for Terzaghi consolidation problems are described briefly at first.

## 2.1. *Problem definition of Terzaghi's consolidation theory*

In Terzaghi consolidation theory, it is assumed that soil particles are incompressible and the dissipation of excess water pressure follows Darcy's law. The volume change of saturated soil is solely attributed to the net flow of pore water. Within a solution domain Ω, the consolidation problem is governed by the following equations, along with the corresponding initial and boundary conditions:

$$\frac{\partial u(\boldsymbol{x},t)}{\partial t} - c\nabla^2 u(\boldsymbol{x},t) = 0, \quad \boldsymbol{x} \in \Omega,\ t \in [0,T] \tag{1}$$

$$u(\boldsymbol{x},0) = u_0(\boldsymbol{x}), \quad \boldsymbol{x} \in \Omega \tag{2}$$

$$\mathcal{B}[u(\boldsymbol{x},t)] = h(\boldsymbol{x},t), \quad \boldsymbol{x} \in \partial\Omega,\ t \in [0,T] \tag{3}$$

where $u(\boldsymbol{x}, t)$ is the excess pore water pressure, which is the function of spatial variable $\boldsymbol{x}$ and time $t$; $c$ denotes the consolidation coefficient and is assumed as a constant; $\nabla^2$ is the Laplace operator; $u_0(\boldsymbol{x})$ is the initial excess water pressure; $\mathcal{B}[\bullet]$ denotes the operator for Dirichlet/Neumann boundary conditions. It is also worth emphasizing that in Eq. (1) the soil is assumed to be isotropic, meaning its vertical and horizontal consolidation coefficients are identical. This study focuses on solving consolidation equations by physics-informed machine learning (PIML) techniques, and therefore, the validity and soundness of these equations are beyond the scope of this work.

## 2.2. *PINN framework for consolidation analyses*

PINNs approximate the exact solutions for PDEs by a neural network architecture, as shown in Figure 1. The spatial and temporal coordinates, $\boldsymbol{x}$ and $t$, are fed into the input layer, and the trial function $u(\boldsymbol{x}, t)$ is obtained after propagating through multiple hidden layers. The predicted $u(\boldsymbol{x}, t)$ is then substituted into the physical laws to compute the residuals, which define the overall loss function as follows:

$$\mathcal{L}(\boldsymbol{\theta}) = \lambda_{\text{pde}}\mathcal{L}_{\text{pde}}(\boldsymbol{\theta}) + \lambda_{\text{bc}}\mathcal{L}_{\text{bc}}(\boldsymbol{\theta}) + \lambda_{\text{ic}}\mathcal{L}_{\text{ic}}(\boldsymbol{\theta}) \tag{4}$$

$$\mathcal{L}_{\text{pde}}(\boldsymbol{\theta}) = \frac{1}{N_c}\sum_{i=1}^{N_c}\left|\frac{\partial u}{\partial t}(\boldsymbol{x}_i, t_i; \boldsymbol{\theta}) - c\nabla^2 u(\boldsymbol{x}_i, t_i; \boldsymbol{\theta})\right|^2 \tag{5}$$

$$\mathcal{L}_{\text{bc}}(\boldsymbol{\theta}) = \frac{1}{N_B}\sum_{j=1}^{N_B}\left|\mathcal{B}\left[u(\boldsymbol{x}_j, t_j)\right] - h(\boldsymbol{x}_j, t_j)\right|^2 \tag{6}$$

$$\mathcal{L}_{\text{ic}}(\boldsymbol{\theta}) = \frac{1}{N_I}\sum_{l=1}^{N_I}\left|u(\boldsymbol{x}_l, 0; \boldsymbol{\theta}) - u_0(\boldsymbol{x}_l)\right|^2 \tag{7}$$

where $\mathcal{L}_{\text{pde}}$, $\mathcal{L}_{\text{bc}}$, and $\mathcal{L}_{\text{ic}}$ are loss terms related to the governing PDEs, initial conditions and boundary conditions, respectively; $\lambda_{\text{pde}}$, $\lambda_{\text{bc}}$ and $\lambda_{\text{ic}}$ are the weights of corresponding loss terms; $\boldsymbol{\theta}$ denotes the network parameters. In Eqs. (5), (6) and (7), $\{\boldsymbol{x}_i, t_i\}_{i=1}^{N_c}$ are the training points in the solution region Ω, $\{\boldsymbol{x}_j, t_j\}_{j=1}^{N_B}$ are the points on the boundaries of the solution region, and $\{\boldsymbol{x}_l, 0\}_{l=1}^{N_I}$ are the training points generated from initial conditions. The involved partial derivatives in the loss function can be calculated using automatic differentiation (AD). Later, the total loss is minimised by gradient descent methods, and the network parameters $\boldsymbol{\theta}$ are updated epoch by epoch until convergence is reached. Finally, the trained neural network can provide an approximate consolidation solution with the input of $(\boldsymbol{x}, t)$.

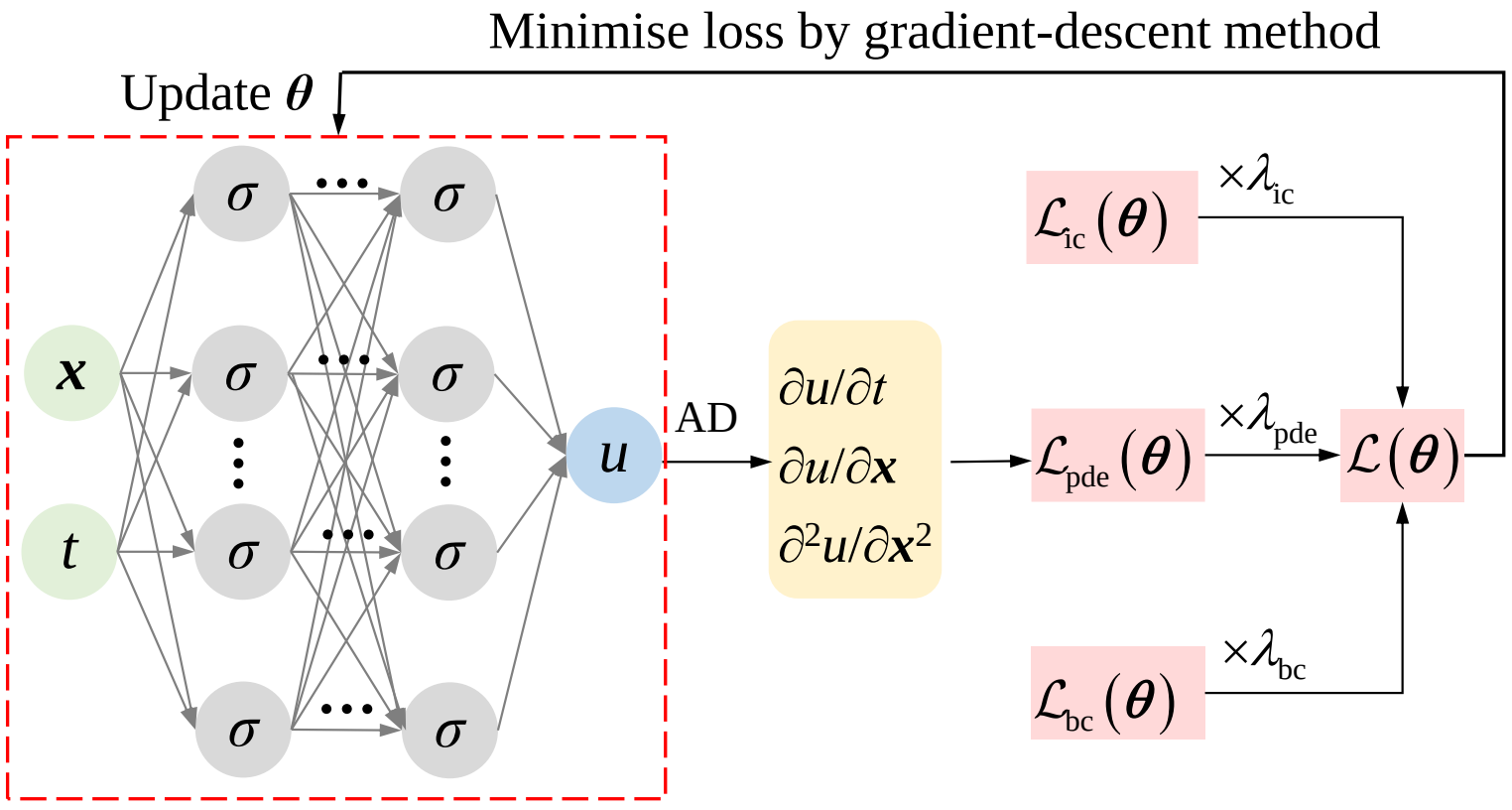

**Figure 1. PINN framework for consolidation analysis**

# 3. PIELM and TS-PIELM frameworks for Consolidation Analyses

## *3.1. Physics-informed extreme learning machine*

In order to abandon the time-consuming gradient descent methods in network training, the PIELM framework finds the latent consolidation solution using ELM instead of fully connected neural networks. As illustrated in Figure 2 (a), there is only one hidden layer in the ELM network with dozens of neurons (the neuron number is denoted as *M*). The input layer parameters are randomly initialised within the range of [-1,1] and remain fixed throughout training. For forward analysis, the governing PDE, initial conditions and boundary conditions are linear, so the output layer parameters are directly determined by minimising the loss vector (physical information at each collocation point), defined as:

$$\left[\mathcal{L}_{\text{pde}}^{(1)}(\boldsymbol{\theta}),\ldots,\mathcal{L}_{\text{pde}}^{(N_c)}(\boldsymbol{\theta}),\mathcal{L}_{\text{bc}}^{(1)}(\boldsymbol{\theta}),\ldots,\mathcal{L}_{\text{bc}}^{(N_B)}(\boldsymbol{\theta}),\mathcal{L}_{\text{ic}}^{(1)}(\boldsymbol{\theta}),\ldots,\mathcal{L}_{\text{ic}}^{(N_I)}(\boldsymbol{\theta})\right]\rightarrow\mathbf{0} \tag{8}$$

$$\mathcal{L}_{\text{pde}}^{(i)}(\boldsymbol{\theta})=\frac{\partial u}{\partial t}(\boldsymbol{x}_i,t_i;\boldsymbol{\theta})-c\nabla^2 u(\boldsymbol{x}_i,t_i;\boldsymbol{\theta}) \quad i=1,2,\ldots,N_c \tag{9}$$

$$\mathcal{L}_{\text{bc}}^{(j)}(\boldsymbol{\theta})=\mathcal{B}\left[u(\boldsymbol{x}_j,t_j)\right]-h(\boldsymbol{x}_j,t_j) \quad j=1,2,\ldots,N_B \tag{10}$$

$$\mathcal{L}_{\text{ic}}^{(l)}(\boldsymbol{\theta})=u(\boldsymbol{x}_l,0;\boldsymbol{\theta})-u_0(\boldsymbol{x}_l) \quad l=1,2,\ldots,N_I \tag{11}$$

where $\boldsymbol{\theta}$ denotes the output layer parameters in ELM networks. In the forward problem, the governing PDE (1), initial conditions (2) and boundary conditions (3) are linear, so $\boldsymbol{\theta}$ can be directly obtained by the least squares method with the Moore–Penrose generalised inverse. Then the solution for Terzaghi consolidation problems can be obtained by the trained ELM. Compared to the PINN employing the time-consuming gradient descent method, the PIELM framework can directly calculate the network parameters to reduce the training cost.

The PIELM network for inverse analysis is illustrated in Figure 2 (b). In contrast to the forward analysis, the coefficient of consolidation is treated as an unknown parameter and must be inferred during the training process. Accordingly, several modifications to the forward PIELM framework are required:

(i) Both the output layer weights $\boldsymbol{\theta}$ and the coefficient of consolidation $c$ are treated as trainable parameters;

(ii) Measured data are necessary and incorporated into the loss vector in the form of $\mathcal{L}_{\text{data}}^{(m)}(\boldsymbol{\theta}) = u(\boldsymbol{x}_m, 0; \boldsymbol{\theta}) - u_{\text{data}}(\boldsymbol{x}_m), m = 1, 2, \ldots, N_D$, where $N_D$ is the total number of measured data.

(iii) Iterative least squares method is adopted to training the ELM network, since the governing PDE (1) becomes nonlinear when coefficient of consolidation is not known in advance.

It should be noted that, when measured data are incorporated into the ELM-based inverse analysis, certain physical constraints can be relaxed or omitted. For CPTu dissipation tests, the initial conditions are not required as long as a sufficient amount of measured data is available.

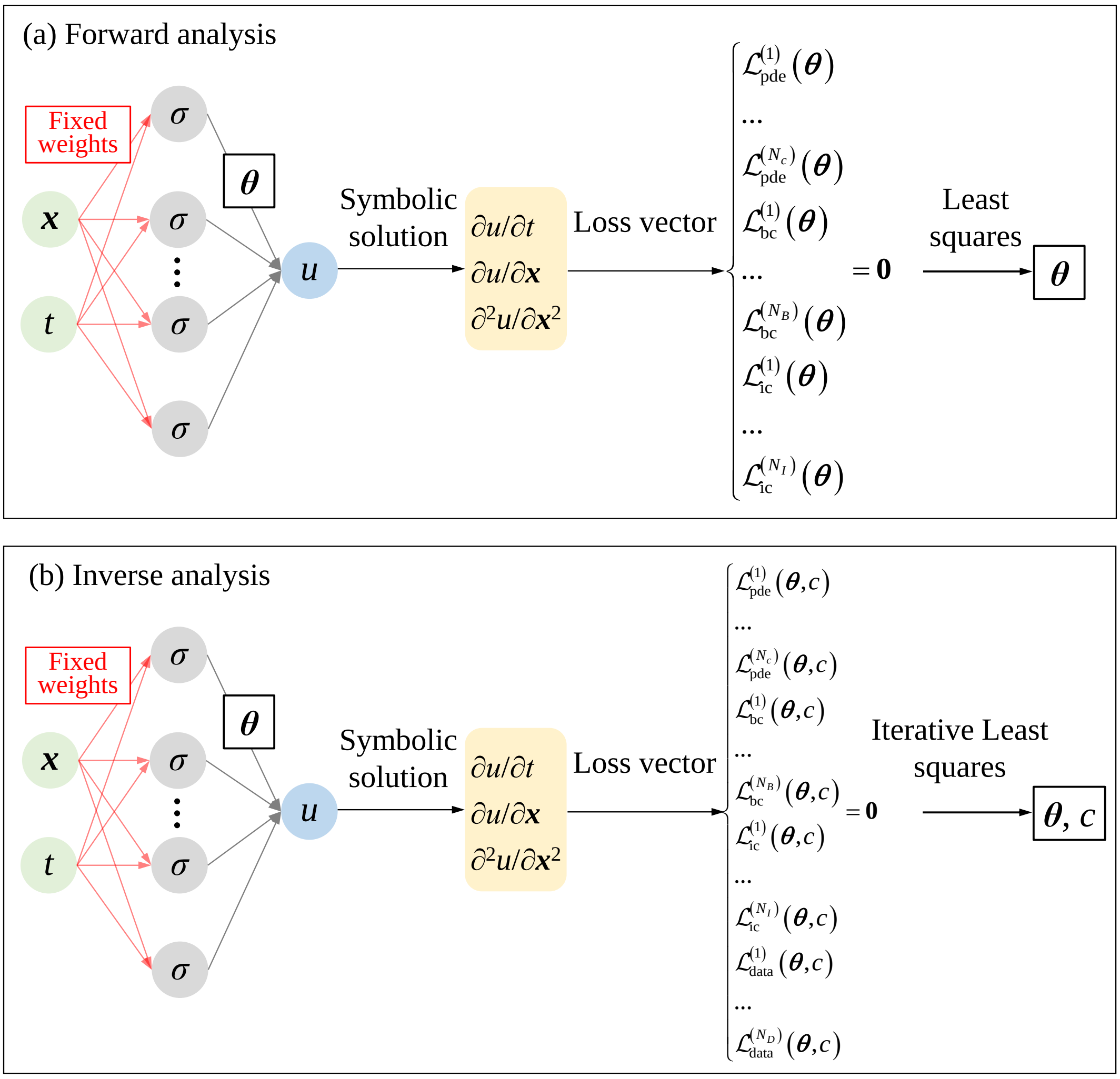

**Figure 2. PIELM framework for consolidation analysis: (a) forward analysis; (b) inverse analysis**

### *3.2. Time-stepping technique*

For forward analysis, when boundary loads are applied suddenly, the prescribed boundary conditions may be inconsistent with the initial distribution of excess pore water pressure at the loaded boundaries. The resulting discontinuity between the initial and boundary conditions can lead to steep pressure gradients over a short dissipation period, which poses challenges for numerical approximation.

Similar to time-stepping physics-informed neural networks (PINNs), a time-stepping strategy can be incorporated into the PIELM framework, leading to the development of the time-stepping PIELM (TS-PIELM). The TS-PIELM framework is presented in this subsection and is primarily intended for forward analysis.

In TS-PIELM framework, the consolidation process is divided into a number of time intervals that can either be of equal or varying lengths. Taking equal-length intervals as an example, the consolidation equation (1) can be written in the Crank–Nicolson discrete form as

$$u_{k+1}(\boldsymbol{x}) = u_k(\boldsymbol{x}) + \frac{\Delta t}{2} c\left[\nabla^2 u_{k+1}(\boldsymbol{x}) + \nabla^2 u_k(\boldsymbol{x})\right] \tag{12}$$

where $\Delta t$ denotes the time interval; $u_k$ denotes the excess water pressure when $t = t_k = k\Delta t$ ($k$=0, 1, 2, …). The TS-PIELM architecture for solving the consolidation equations is shown in Figure 3. Different from the standard PIELM framework, a separate ELM network is required at each time step. The output $u_k$ from the $k$-th ELM network will serve as the initial conditions for the ($k$+1)-th ELM network. Letting $u_k(\boldsymbol{x};\boldsymbol{\theta}_k) = u(\boldsymbol{x}, t_k)$ and $\boldsymbol{\theta}_k$ denotes the weight parameters of $k$-th network, the loss vector at the ($k$+1)-th incremental step will be

$$\boldsymbol{\mathcal{L}}(\boldsymbol{\theta}_{k+1}) = \left[\mathcal{L}_{\text{pde}}^{(1)}(\boldsymbol{\theta}_{k+1}), \ldots, \mathcal{L}_{\text{pde}}^{(N_c)}(\boldsymbol{\theta}_{k+1}), \mathcal{L}_{\text{bc}}^{(1)}(\boldsymbol{\theta}_{k+1}), \ldots, \mathcal{L}_{\text{bc}}^{(N_b)}(\boldsymbol{\theta}_{k+1})\right] \to \mathbf{0} \tag{13}$$

$$\mathcal{L}_{\text{pde}}^{(i)}(\boldsymbol{\theta}_{k+1}) = u_{k+1}(\boldsymbol{x}_i;\boldsymbol{\theta}_{k+1}) - \frac{\Delta t}{2} c\left[\nabla^2 u_{k+1}(\boldsymbol{x}_i;\boldsymbol{\theta}_{k+1}) + \nabla^2 u_k(\boldsymbol{x}_i;\boldsymbol{\theta}_k)\right] - u_k(\boldsymbol{x}_i;\boldsymbol{\theta}_k) \tag{14}$$

$$\mathcal{L}_{\text{bc}}^{(i)}(\boldsymbol{\theta}_{k+1}) = \mathcal{B}\left[u_{k+1}(\boldsymbol{x}_j;\boldsymbol{\theta}_{k+1})\right] - h(\boldsymbol{x}_j) \tag{15}$$

Following a similar procedure of PIELM in Section 3.1, $\boldsymbol{\theta}_{k+1}$ can be readily calculated by optimising $\boldsymbol{\mathcal{L}}(\boldsymbol{\theta}_{k+1})=0$. Subsequently, $u_{k+1}(\boldsymbol{x};\boldsymbol{\theta}_{k+1})$ can be determined by the trained ELM network, and $u(\boldsymbol{x}, t)$ can be computed step by step with increasing $k$. It can be found that the TS-PIELM requires computing the Moore-Penrose inverse at each time step rather than over the whole dissipation period. In the next sections, we will demonstrate that the time-stepping technique is helpful to solve the consolidation equation when the initial conditions and boundary conditions are discontinuous. However, since the whole consolidation process is separated into time-intervals, TS-PIELM encounters difficulties in parameter identification for inverse problems.

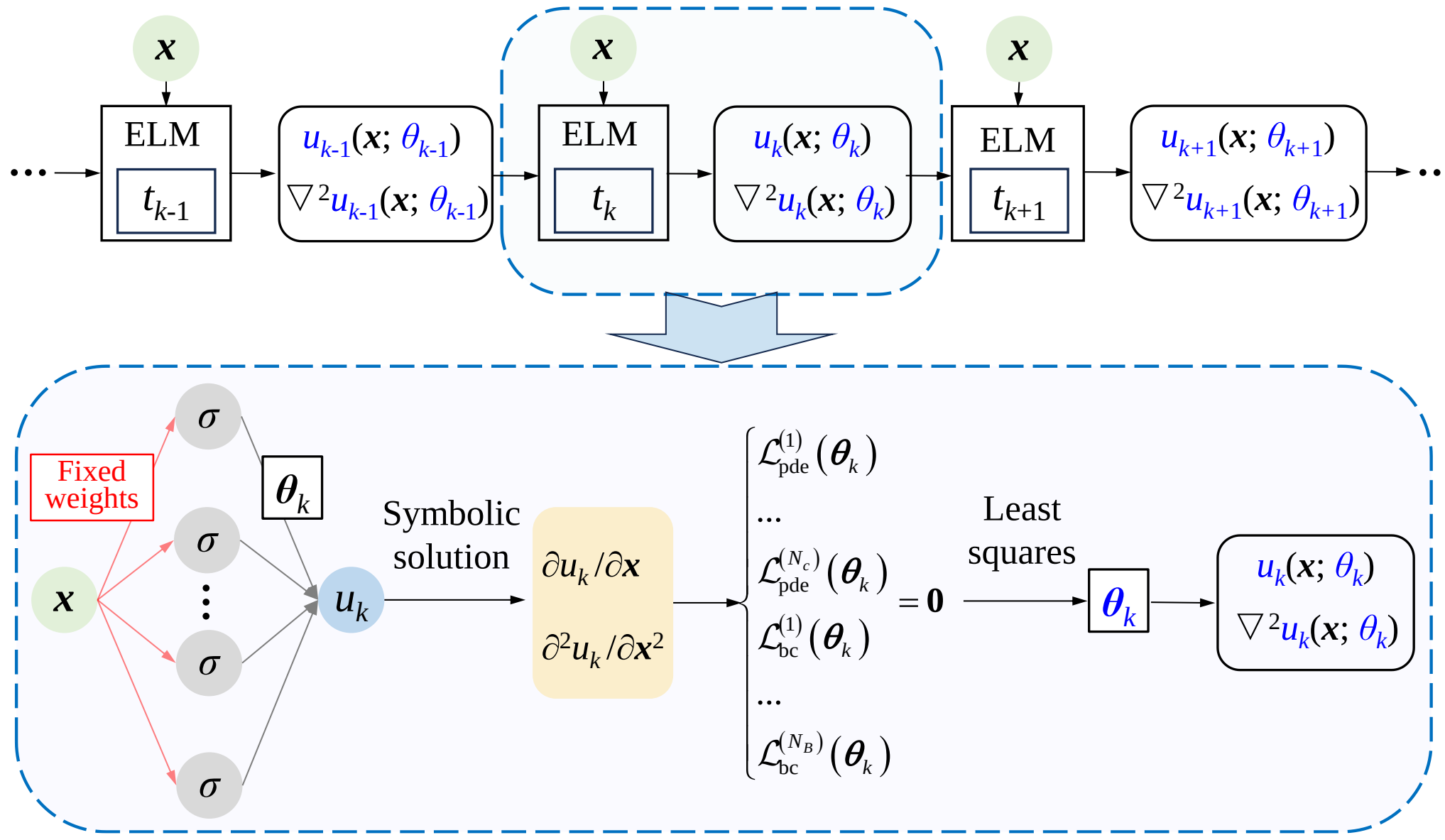

**Figure 3. TS-PIELM framework for consolidation analysis**

## 4. Forward analysis: validation

The PIELM method for Terzhaghi consolidation problems are first validated by forward problems. For comparison, results obtained from PINN are also presented. All simulations are conducted using MATLAB R2023b on a Dell Precession 7960 workstation, equipped with an Intel Xeon W7-3445 (2.59 GHz) processor and 64GB of RAM. The hyperparameters of the neural networks as well as training details for the three cases are summarised in Table 1. Following Lu et al. (2021), Schiassi et al. (2021), and Lan et al. (2024a), hard constraints are imposed on the governing equations and boundary conditions, and the corresponding expressions are provided in the Appendix. Furthermore, the self-adaptive loss balancing technique based on maximum likelihood estimation is introduced to dynamically update loss weights at each training epoch (Xiang et al. 2022). To assess the predictive accuracy, the relative $L_2$ error is defined as follows:

$$L_2 = \frac{\sqrt{\sum_1^N (\text{exact -predicted})^2}}{\sqrt{\sum_1^N (\text{exact})^2}} \tag{16}$$

where $N$ is the number of test points.

**Table 1 Summary of hyperparameters and training information**

| Cases | Case 1 | | | Case 2 | | | Case 3 | | |
|---|---|---|---|---|---|---|---|---|---|
| Frameworks | PIELM | PINN | TS-PIELM | PIELM | PINN | TS-PIELM | PIELM | PINN | TS-PIELM |
| Neurons in hidden layers | 100 | 64*3 | 20 | 100 | 64*3 | 20 | 1000 | 64*3 | 150 |
| Activation function | tanh | tanh | tanh | tanh | tanh | tanh | tanh | tanh | tanh |
| $N_c$ | 2000 | 2000 | 1000 | 2000 | 2000 | 1000 | 20000 | 20000 | 10000 |
| $N_I$ | 100 | 100 | - | 100 | 100 | - | 2000 | 2000 | |
| Time-interval (s) | - | - | 0.001 | | | 0.001 | | | 0.01 |
| Training time (s) | 0.86 | 434.10 | **0.37** | 0.99 | 405.41 | **0.34** | 47.82 | 5231.85 | **10.14** |
| $L_2$ | 5.07e-2 | 3.28e-2 | **2.94e-04** | 1.48e-2 | 0.0213 | **4.90e-04** | 9.32e-2 | 9.43e-2 | **2.50e-3** |

Note: For PINN, epoch=5000; initial learning rate=0.001; declay rate=0.005

### *4.1. One-dimensional consolidation in Certain coordinates*

The PIELM approach is first validated by a one-dimensional vertical consolidation problem, which is a well-established benchmark for assessing the performance of PINN in consolidation analysis. Within the solution domain of $0<z<H$, the top boundary ($z$=0) is a drainage outlet and the bottom boundary ($z$=$H$) is impermeable. The excess pore water pressure $u(z,t)$ is governed by

$$\begin{cases} \dfrac{\partial u}{\partial t} = c\dfrac{\partial^2 u}{\partial z^2} \\ u(z,0) = p_0 \\ u(0,t) = 0 \\ \left.\dfrac{\partial u}{\partial z}\right|_{z=H} = 0 \end{cases} \tag{17}$$

where $p_0$ is the initial uniform excess water pressure; $H$ is the vertical distance from the top to the bottom boundary. The exact solution of Eq. (17) and the dimensionless time ($T$) are expressed as

$$\frac{u(z,t)}{p_0} = \frac{4}{\pi}\sum_{n=1}^{\infty}\frac{1}{(2n+1)}\sin\left[\frac{(2n+1)\pi z}{2H}\right]\exp\left[-\frac{(2n+1)^2\pi^2 ct}{4H^2}\right] \tag{18}$$

$$T = ct/H^2 \tag{19}$$

Figure 4 compares the predicted excess pore water pressure $u(z,t)$ obtained using the PIELM and PINN approaches with the exact solution at different consolidation times (i.e. $T$=0.05, 0.1, 0.2, 0.5, 1). Figure 4 shows that both PIELM and PINN predictions are in good agreement with the exact solution, yielding relative $L_2$ error of 5.92e-2 and 3.28e-2, respectively. In terms of computational efficiency, PINN requires a significantly longer training time (approximately 434 s), which can be attributed to the inherent limitations of deep neural networks. In contrast, the training time of PIELM is drastically reduced to 0.86s owing to the adoption of the extreme learning machine (ELM) architecture. The prediction errors mainly originate from the discontinuity between the initial and boundary conditions at short consolidation times (e.g., $T$<0.05).

Previous studies have addressed this issue by introducing continuous boundary conditions following Mei and Lok (2014) within PINN frameworks, achieving relative $L_2$ errors on the order of 1e-4 (Lan et al. 2024a; Zhou et al. 2025).

To highlight the difficulty induced by the discontinuity considered in this study, contour plots of the absolute point-wise errors (i.e., absolute value of the predicted minus the exact $u/p_0$) are presented in Figure 5. The results obtained using the TS-PIELM framework are also included for comparison. It can be observed that:

(a) For both PIELM and PINN, the absolute errors remain at relatively high levels at very early times and gradually decrease to the order of 1e-4 when $T$>0.05. The relative $L_2$ error is primarily influenced by the predictions at very small times ($T$<0.01). This behavior arises because, when $t$ approaches zero, the initial conditions are incompatible with the boundary conditions at $z$=0, resulting in excessively steep gradients of $u(z, t)$ and consequently larger numerical errors.

(b) For TS-PIELM, the absolute errors range from 1e-5 to 1e-2, but rapidly decrease to below 1e-5 once $T$>0.01. This observation demonstrates that the time-stepping technique effectively smooths the steep gradients induced by discontinuities and significantly improves the prediction accuracy at early consolidation times. The training time of TS-PIELM is also shorter than that of vanilla PIELM. This is primarily because fewer training points and hidden layer neurons are required in each time interval, which reduces memory usage and improves computational efficiency.

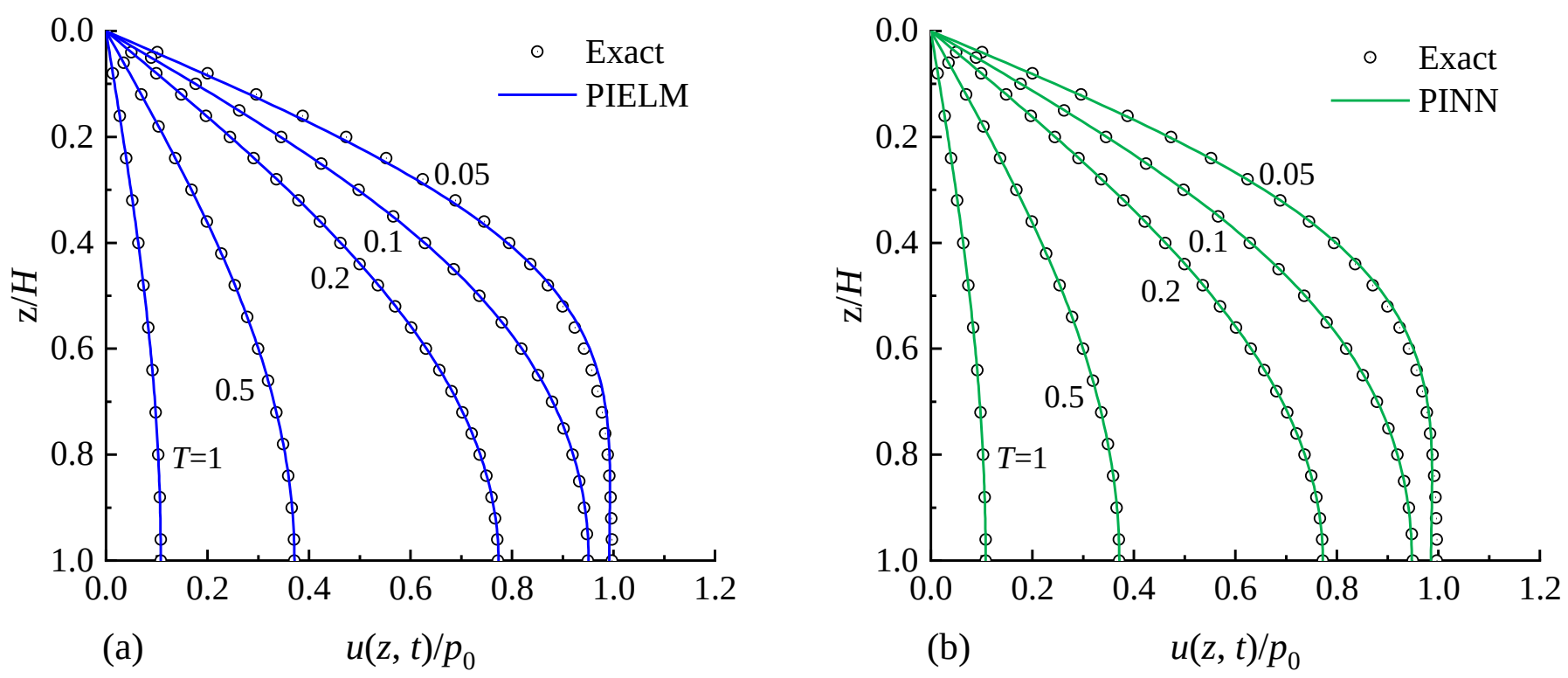

**Figure 4. Excess pore water pressures predicted by PIELM and PINN against the exact solution**

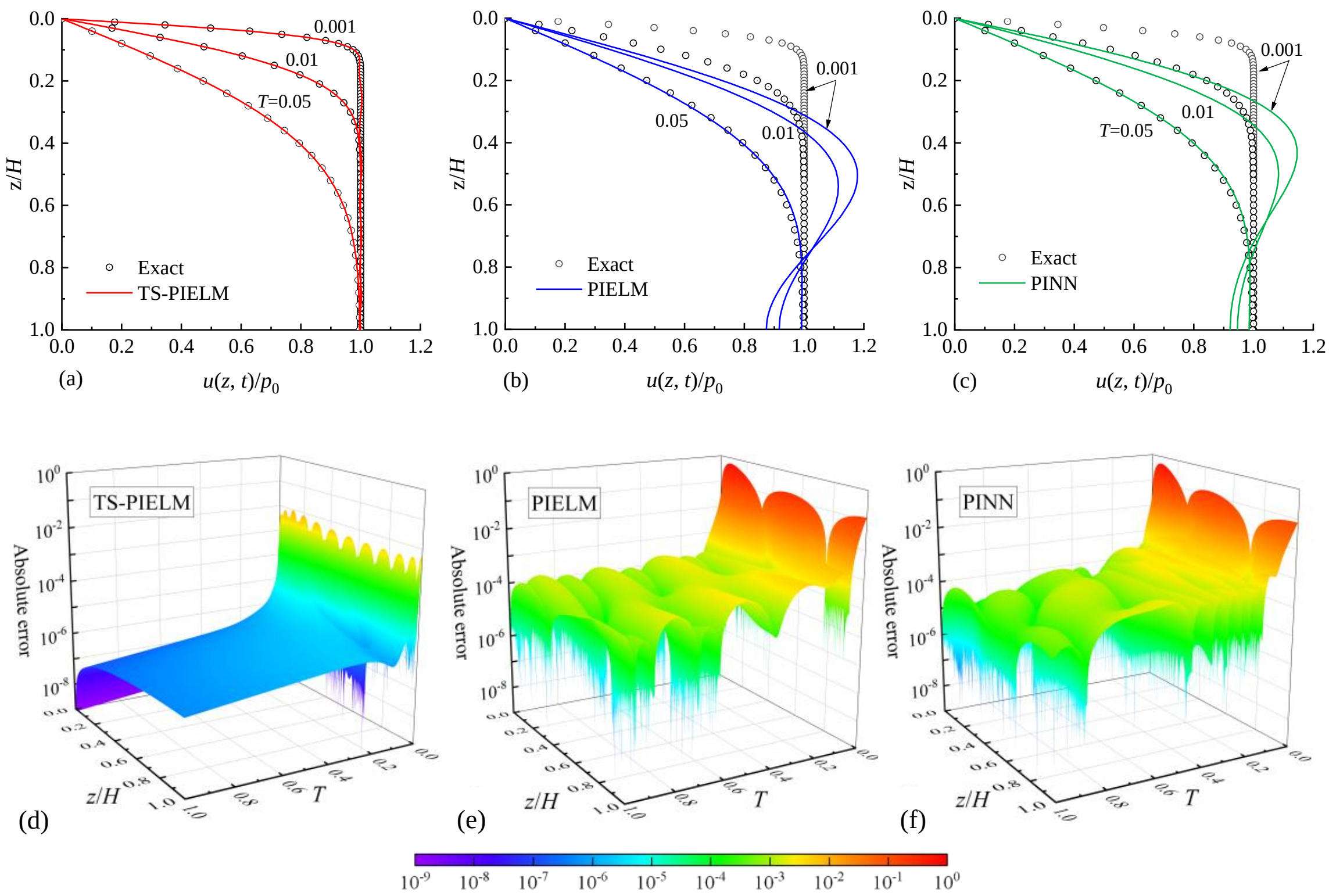

**Figure 5. Excess pore water pressures and absolute error contours for TS-PIELM, PIELM and PINN**

### 4.2. *One-dimensional radial consolidation*

The second illustrative case is a one-dimensional radial consolidation problem defined within the domain range of $a<r<b$, where the boundary $r=a$ is permeable and the boundary $r=b$ is subjected to constant pressure $p_0$ representing a far zone. In this case, $u(r,t)$ is governed by

$$\begin{cases} \dfrac{\partial u}{\partial t} = c\left(\dfrac{\partial^2 u}{\partial r^2} + \dfrac{1}{r}\dfrac{\partial u}{\partial r}\right) \\ u(r,0) = p_0 \\ u(a,t) = 0 \\ u(b,t) = p_0 \end{cases} \tag{20}$$

The analytical solution can be found in Carslaw and Jaeger (1947) and Yang et al. (2024), and it can be expressed as

$$\frac{u}{p_0}=1-\frac{\ln(b/r)}{\ln(b/a)}-\pi\sum_{n=1}^{\infty}\frac{J_0(\beta_n a)J_0(\beta_n b)U_0(\beta_n r)}{J_0^2(\beta_n a)-J_0^2(\beta_n b)}\exp\left(-c\beta_n^2 t\right) \tag{21}$$

$$U_0(\beta_n r)=J_0(\beta_n r)Y_0(\beta_n b)-J_0(\beta_n b)Y_0(\beta_n r) \tag{22}$$

where $J_0$ and $Y_0$ are the first and second kinds of Bessel functions of zero order, respectively; $\alpha_n$ is the $n$-th root of $U_0(\alpha_n a)=0$. For this case, the dimensionless time is defined as $T$=$ct/a^2$.

Figure 6 shows the comparison of $u(r,t)$ between the exact solution and predicted results by the three PIML approaches, plotted on a semi-logarithmic scale. Once again, PIELM can provide acceptable predictions when the consolidation time is longer, and requires much lower training time than PINN. For a short consolidation period, both PIELM and PINN fail, yet TS-PIELM demonstrates an excellent agreement with the exact solution when $T$<0.01.

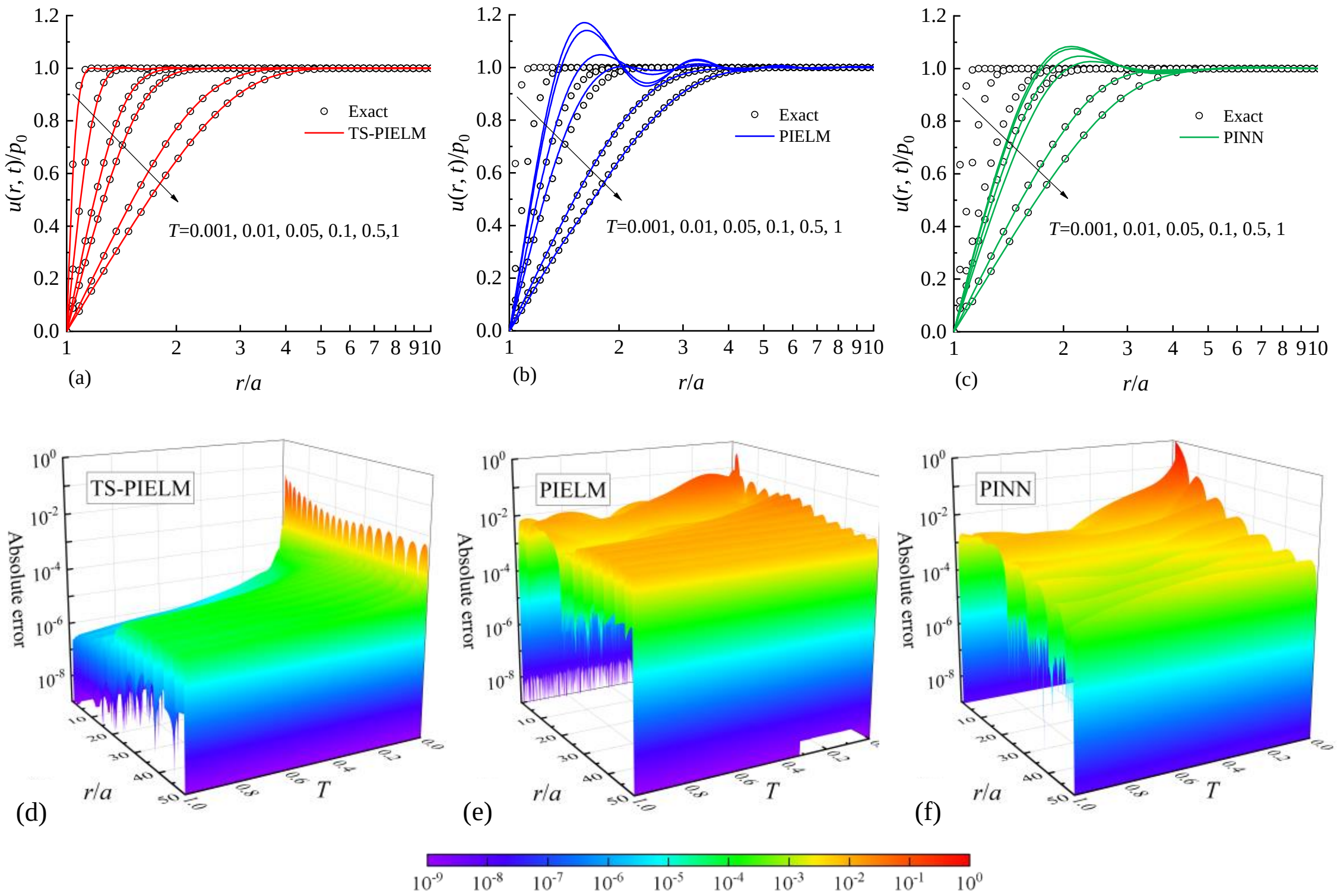

**Figure 6. Excess pore water pressures and absolute error contours predicted by TS-PIELM, PIELM and PINN**

### *4.3. Two-dimensional radial consolidation with vertical drains*

The final case study examines two-dimensional radial consolidation with prefabricated vertical drains (PVDs), a widely adopted technique to enhance drainage and accelerate soft clay consolidation (Indraratna et al. 2009; Geng and Yu 2017; Ni et al. 2022). As shown in Figure 7, within the solution domain of $a<r<b$ and $0<z<H$, the problem is defined by Eq. (23):

$$\begin{cases} \dfrac{\partial u}{\partial t} = c\left(\dfrac{\partial^2 u}{\partial r^2} + \dfrac{1}{r}\dfrac{\partial u}{\partial r} + \dfrac{\partial^2 u}{\partial z^2}\right) \\ u(r,z,0) = p_0 \\ u(a,z,t) = 0 \\ \left.\dfrac{\partial u}{\partial r}\right|_{r=b} = 0 \\ u(r,0,t) = 0 \\ \left.\dfrac{\partial u}{\partial z}\right|_{z=H} = 0 \end{cases} \tag{23}$$

Using the method of separation of variables (Carslaw and Jaeger 1947), the analytical solution of Eq. (23) can be expressed as

$$\frac{u(r,z,t)}{p_0} = F_r F_z \tag{24}$$

$$F_z = \frac{4}{\pi}\sum_{n=1}^{\infty}\frac{1}{(2n+1)}\sin\left(\frac{(2n+1)\pi z}{2H}\right)\exp\left(-\frac{(2n+1)^2\pi^2}{4}\frac{ct}{H^2}\right) \tag{25}$$

$$F_r = \sum_{n=1}^{\infty}\frac{\int_a^b U(r)r\mathrm{d}r}{\int_a^b U^2(r)r\mathrm{d}r}U(r)\exp\left(-c\beta_n{}^2 t\right) \tag{26}$$

$$U(r) = Y_0(\beta_n r) - \frac{Y_0(\beta_n a)}{J_0(\beta_n a)}J_0(\beta_n r) \tag{27}$$

where $\beta_n$ is the n-th root of the $J_0(\beta a)Y_1(\beta b) - Y_0(\beta a)J_1(\beta b) = 0$, where $Y_1$ and $J_1$

are the first and second kinds of Bessel functions of first order, respectively.

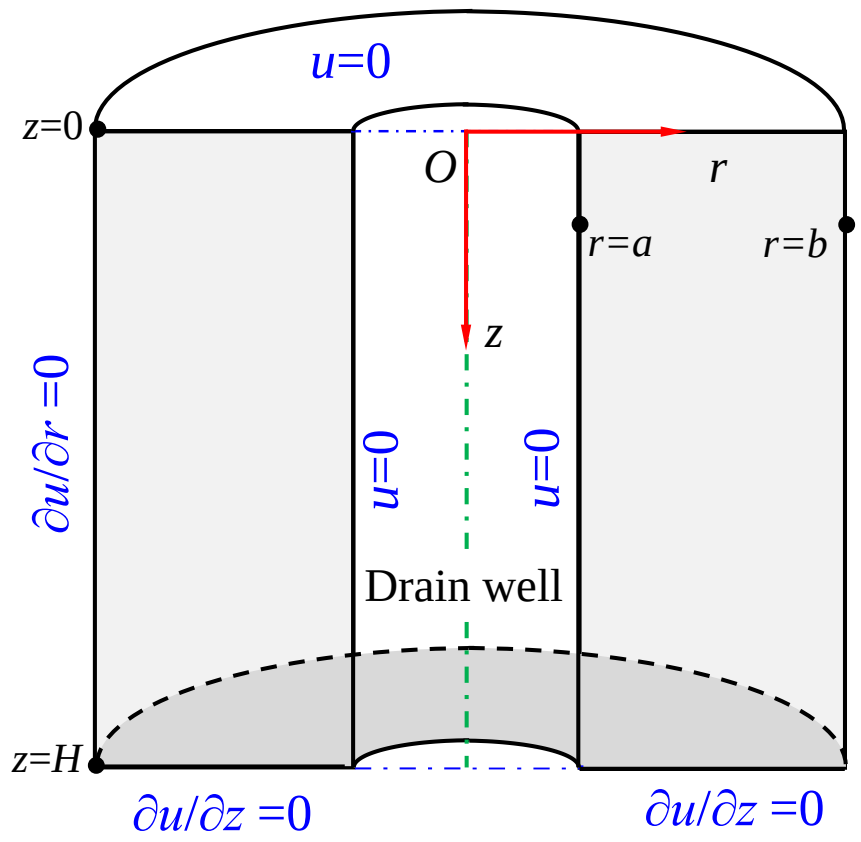

**Figure 7. Schematic of radial consolidation with prefabricated vertical drain**

In this case the PVD sizes are set as $a$=0.07m and $b$=0.7m (Geng and Yu 2017), and the dimensionless time is defined as $T = ct/H^2$ where $H$=5m. Figure 8 and Figure 9 show the predicted $u(r, z,t)/p_0$ and absolute point-wise error on different occasions, respectively. Owing to more training points and hidden layer neurons required in ELM networks, the training time for TS-PIELM and PIELM increased to 10s and 48s, respectively, which is significantly lower than that of PINN after 5000 epochs (over 1 hour). As shown by the contour plots of $u/p_0$ and absolute error, TS-PIELM can accurately predict the pore water pressure evolution in this two-dimensional consolidation problem with discontinuous initial and boundary conditions at two boundaries ($z$=0 and $r$=$a$), giving the relative $L_2$ error of 2.50e-3. PIELM and PINN perform worse in terms of both relative $L_2$ error and absolute error, especially when $T$<4e-3 (i.e., $ct$<0.1m$^2$). As aforementioned, however, TS-PIELM framework is not suitable for inverse analysis, so in this section the results predicted by TS-PIELM are discussed mainly for the purpose of validation.

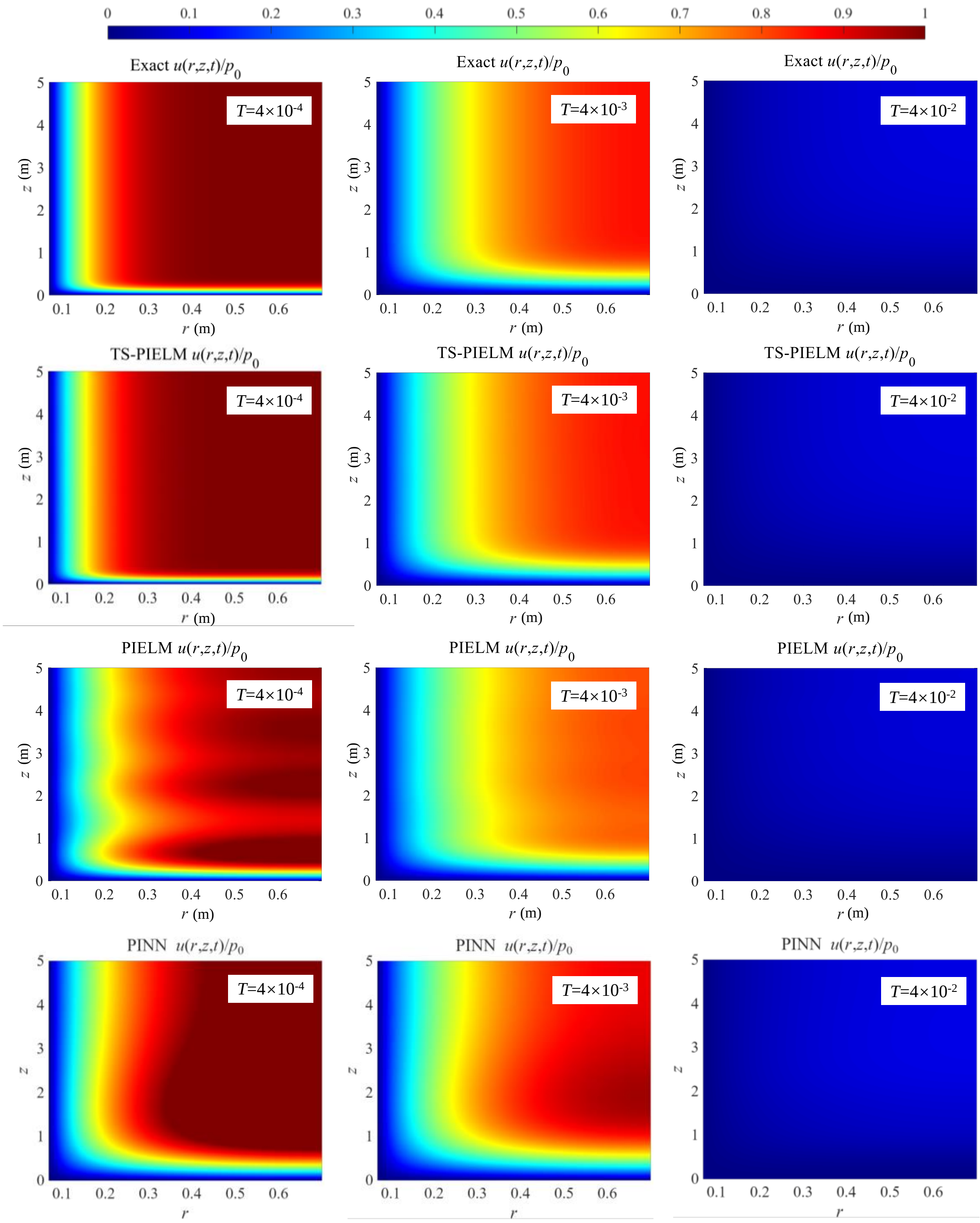

**Figure 8. Predicted $u/p_0$ by TS-PIELM, PIELM and PINN**

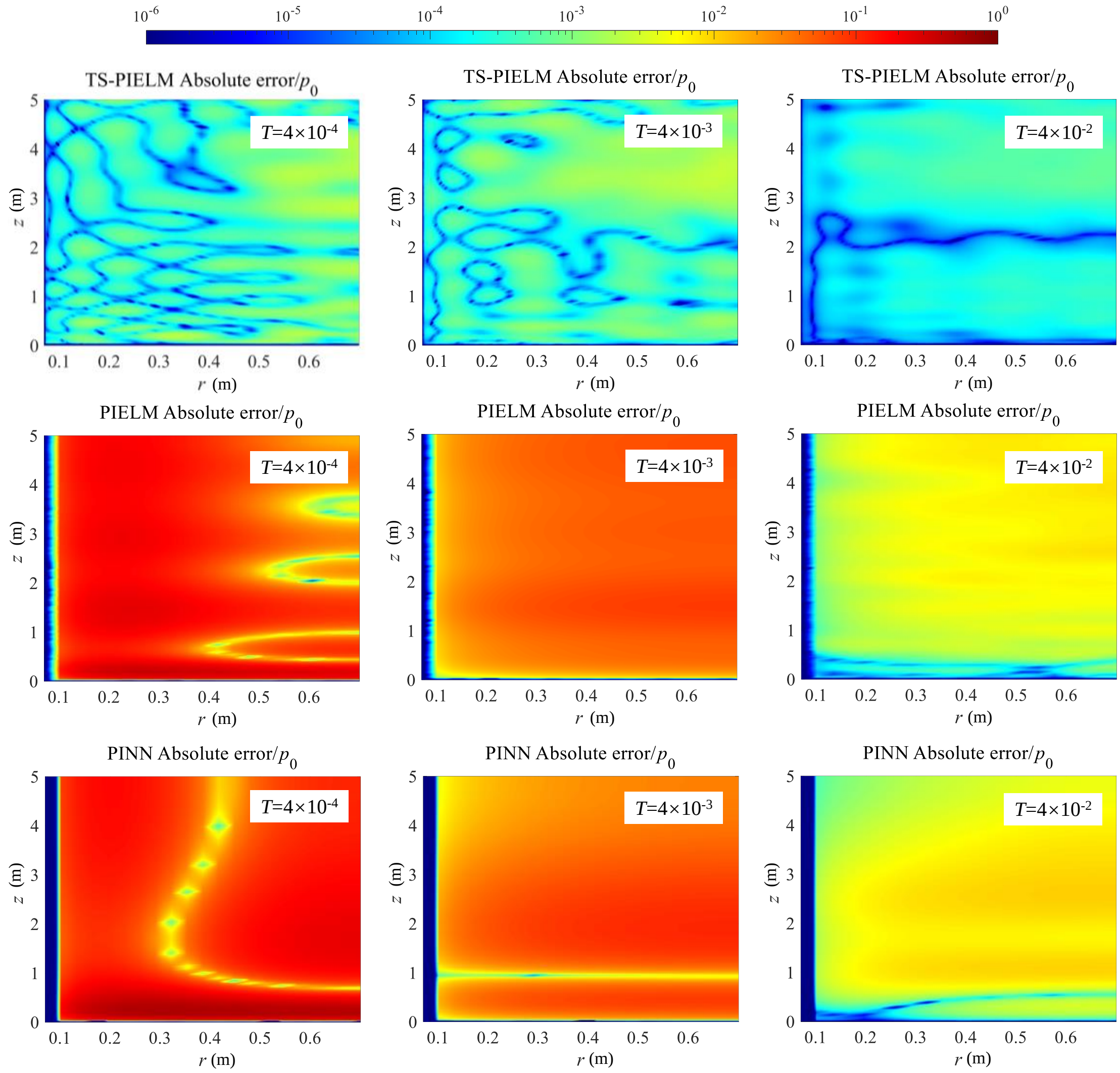

**Figure 9. Absolute error of $u/p_0$ for TS-PIELM, PIELM and PINN**

### 4.4. *Parametric studies for forward analysis*

After validating the PIELM approach, parametric studies are conducted to investigate the robustness of the method, as well as the effects of the number of hidden neurons, the number of collocation points, and the choice of time intervals in TS-PIELM. Unless stated otherwise, Case 1 in Section 4.1 is used for illustration.

First of all, the robustness of PIELM and TS-PIELM is examined, as the input weights in the ELM network are randomly assigned. Table 2 presents the relative $L_2$ errors and training times for five repeated tests using the same parameters as in Table 1. For both PIELM and TS-PIELM, the results are observed to be stable, exhibiting low variance across repeated runs.

**Table 2 Repeated numerical tests on the performance of PIELM**

| Test number | PIELM | | TS-PIELM | |
|---|---|---|---|---|
| | $L_2$ | Training time | $L_2$ | Training time |
| 1 | 5.11e-02 | 0.86 | 2.91e-04 | 0.37 |
| 2 | 5.07e-02 | 0.75 | 3.60e-04 | 0.36 |
| 3 | 7.98e-02 | 0.74 | 2.90e-04 | 0.37 |
| 4 | 5.08e-02 | 0.78 | 3.00e-04 | 0.37 |
| 5 | 5.13e-02 | 0.78 | 4.19e-04 | 0.35 |
| Mean | 5.67e-02 | 0.78 | 3.32e-04 | 0.36 |
| Variance | 1.32e-04 | 1.8e-03 | 2.56e-09 | 8.93e-05 |

Secondly, parametric studies are conducted to investigate the influence of the number of hidden neurons and collocation points on the prediction efficiency and accuracy of PIELM. Figure 10 and Figure 11 show the training time and relative $L_2$ error for various combinations of collocation points and hidden neurons. With the given parameters, the following observations can be made:

(i) The training time increases almost linearly with the number of hidden neurons. The number of collocation points has little effect on training time unless it is very large (e.g., exceeding 10,000).

(ii) The prediction accuracy initially decreases sharply with increasing $M$, then gradually approaches a steady state. For a given $M$, there exists an optimal number of collocation points that minimizes the relative $L_2$ error.

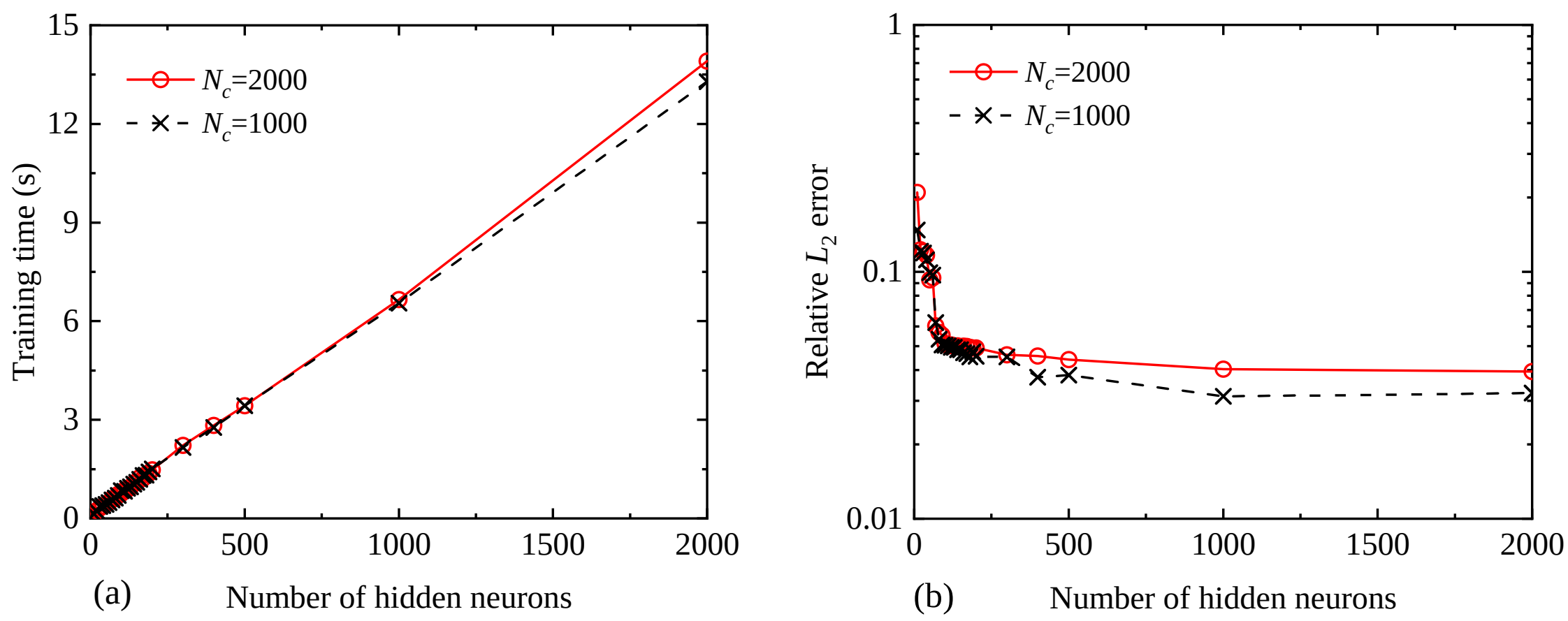

**Figure 10 Influence of hidden neuron numbers on the prection efficiency and accuracy of PIELM: (a) Training time; (b) Relative $L_2$ error.**

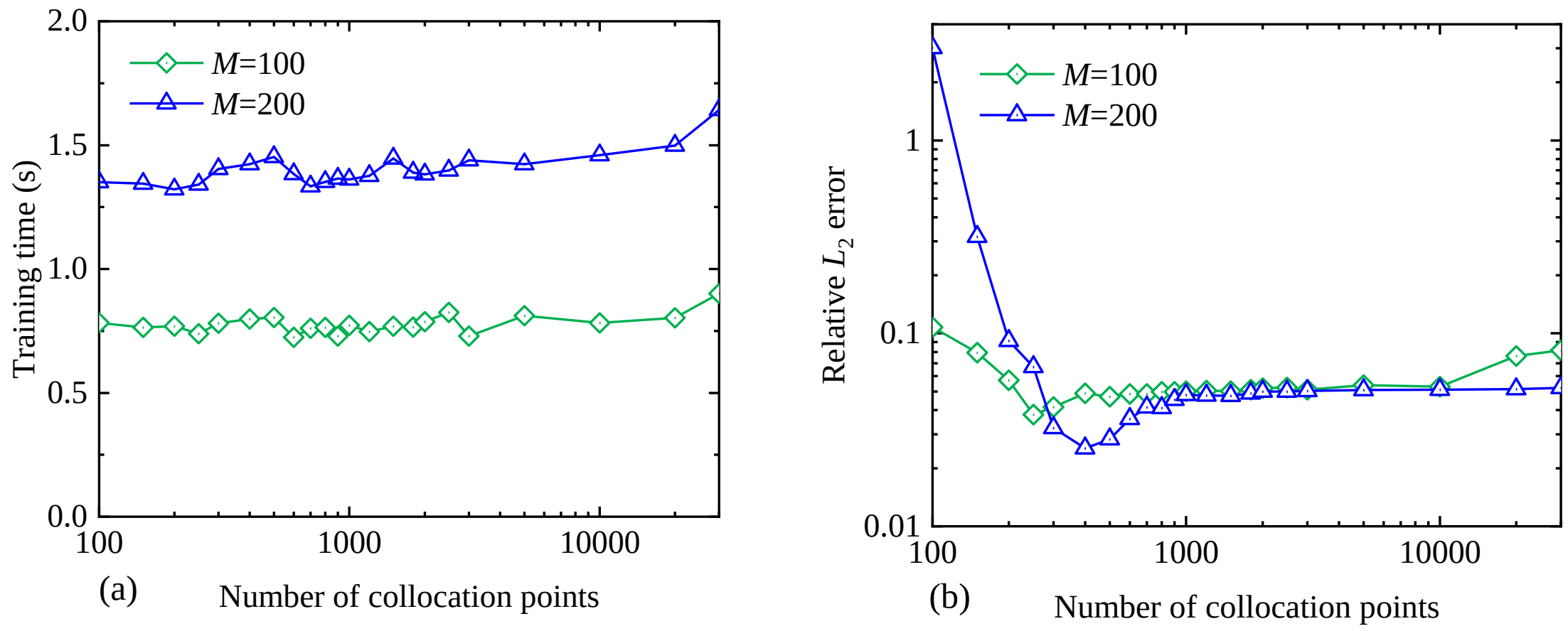

**Figure 11 Influence of collocation point numbers on the prection efficiency and accuracy of PIELM: (a) Training time; (b) Relative $L_2$ error.**

Finally, Figure 12 illustrates the effect of step size ($t_k$-$t_{k-1}$) on the performance of TS-PIELM, using the input parameters listed in Table 1. As the step size decreases (i.e., as the number of time intervals increases), the training time grows and reaches approximately 2s when the time interval is 1e-4s. However, increasing the number of time intervals does not consistently reduce the relative $L_2$ error. In fact, a larger relative $L_2$ error is observed for very small step sizes (e.g., 1e-4 s), and this may be caused by the accumulation of numerical errors over the sequence of fine time steps.

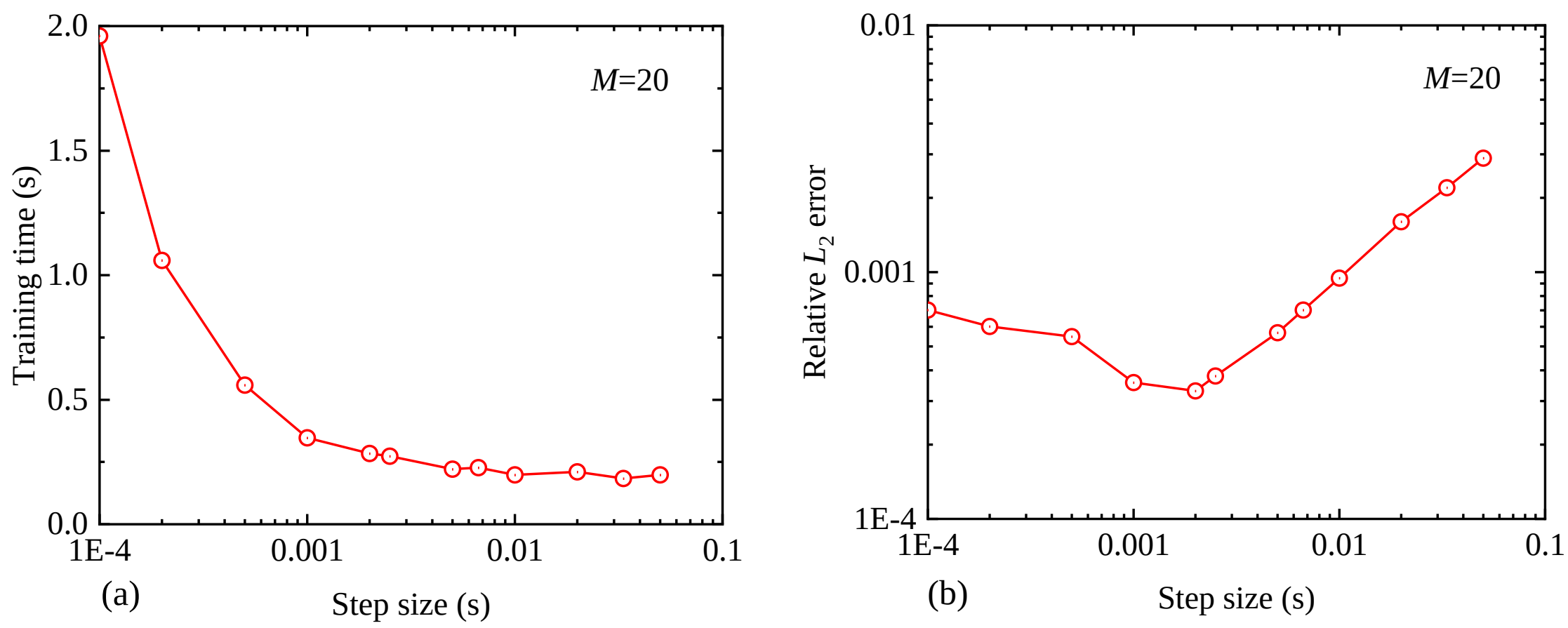

**Figure 12 Influence of step size on the prection efficiency and accuracy of TS-PIELM: (a) Training time; (b) Relative $L_2$ error.**

## 5. Inverse analysis: application to CPTu

One of the main advantages of physics-informed machine learning lies in its capability to address inverse problems. In this section, the application of PIELM to infer the coefficient of consolidation from CPTu dissipation tests is investigated.

The governing partial differential equation for the excess pore water pressure in dissipation tests is typically expressed as

$$\begin{cases} \dfrac{\partial u}{\partial t} = c\left(\dfrac{\partial^2 u}{\partial r^2} + \dfrac{1}{r}\dfrac{\partial u}{\partial r}\right) \\ \partial u/\partial r\big|_{r=D/2} = 0 \\ u(\infty, t) = 0 \\ u(r,0) = u_0(r) \end{cases} \tag{28}$$

where $D$ is the diameter of the piezocone probe; in Eq. (28) the boundary at $r=D/2$ is impermeable, while the excess water pressure is set as zero at a far zone. In this paper the generation and dissipation of excess water pressure are analysed focusing on a horizontal cross section, and vertical water flow is not considered in Eq. (28). As discussed above, the initial conditions $u_0(r)$ for Eq. (28) depend on the CPTu penetration stage and vary from case to case. While it may be difficult to kwow the initial conditions $u_0(r)$ around the probe, the dissipation of excess pore water pressure at $r=D/2$ can be measured, providing valuable information regarding soil permeability. Consequently, the interpretation of CPTu dissipation tests constitutes an inverse problem with partially known physics and partially observed data. Compared with conventional empirical or trial-and-error methods, physics-informed machine learning offers a promising alternative. To evaluate the performance of PIELM for inverse problems, a well-known solution of Randolph and Wroth (1979) is set as a benchmark case. In this case, the initial conditions for the CPTu dissipation test are derived from a simple cavity expansion solution:

$$u_0(r)=\begin{cases}2s_u\ln(\rho/r), & a<r<\rho\\ 0, & r>\rho\end{cases} \tag{29}$$

where $s_u$ is a material constant for the undrained shear strength of soft soils; $\rho = D\sqrt{G/s_u}/2$ is the outer radius of plastic zone, in which $G$ is the shear modulus of soils and can also be regarded as the constant in the dissipation stage. Using the initial conditions (29), an analytical solution was obtained in Randolph and Wroth (1979). This solution is used to generate the "measured" excess pore water pressure at $r=a$ in the CPTu dissipation test. The coefficient of consolidation of the soil is then inferred by PIELM using these measured data. Figure 13 illustrates an example of the interpretation results.

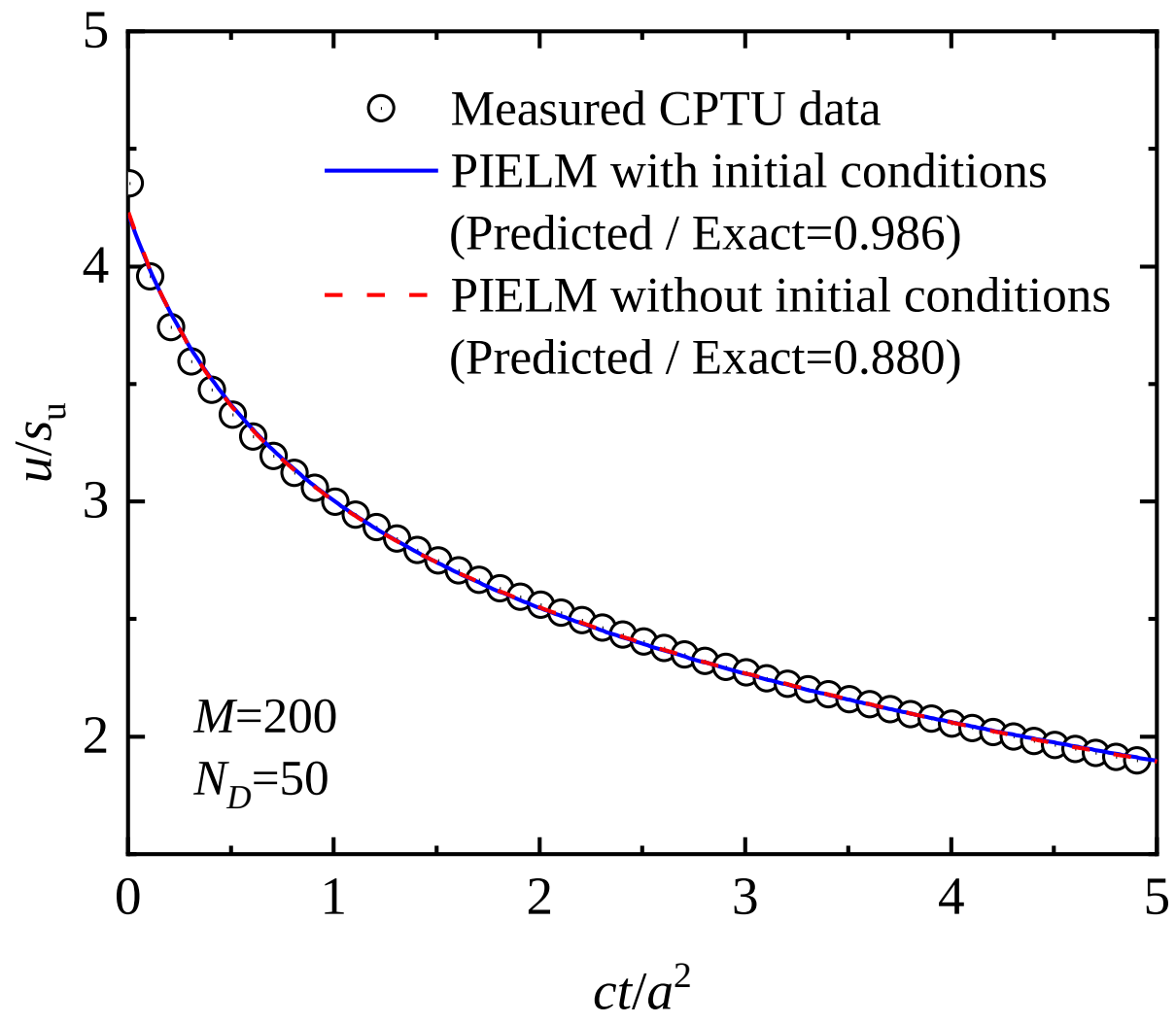

**Figure 13 Example interpretation of CPTu dissipation tests by PIELM with and without initial conditions**

Figure 14 presents the coefficient of consolidation of soil inferred from the "measured" CPTu data using PIELM, with results obtained both with and without prescribed initial conditions for comparison. The numbers of training points are summarized in Table 3. Results are computed from five repeated tests, where the solid lines denote the mean values and the shaded areas represent the corresponding error bands. When the initial conditions are assumed known a priori, the predicted coefficient of consolidation gradually approaches the exact value as the amount of measured CPTu data increases. Comparing Figure 14 (a), (b) and (c), it can also be observed that predictions exhibit higher variability when fewer hidden neurons are used. Moreover, even without prior knowledge of the initial conditions, the coefficient of consolidation can still be inferred provided that sufficient measured data from CPTu dissipation tests are available. The trade-off is that the prediction accuracy is generally lower than that obtained with known initial conditions, and greater variability is observed. In general, when the number of measured data points exceeds approximately 40 and $M$=200 (or 300), the predicted error for the coefficient of consolidation remains within 30%. Considering that the in-situ coefficient of consolidation often varies over a wide range, this level of error is considered acceptable in practice when initial conditions are not known a priori.

**Table 3 Number of training data in inverse problems**

| Collocation points | Hidden neurons | Initial conditions | Boundary conditions | CPTu data |
|---|---|---|---|---|
| 5000 | 100/200/300 | 100 | 100 | 1~500 |

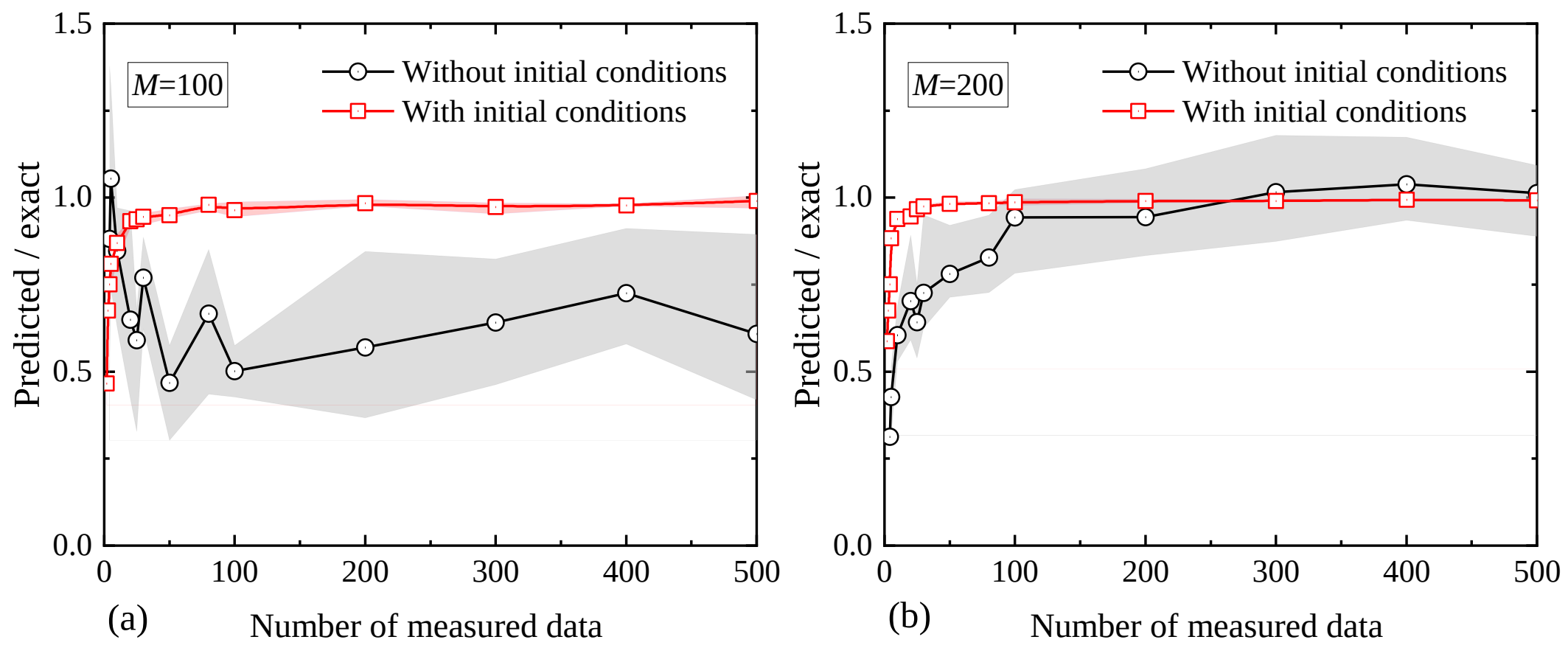

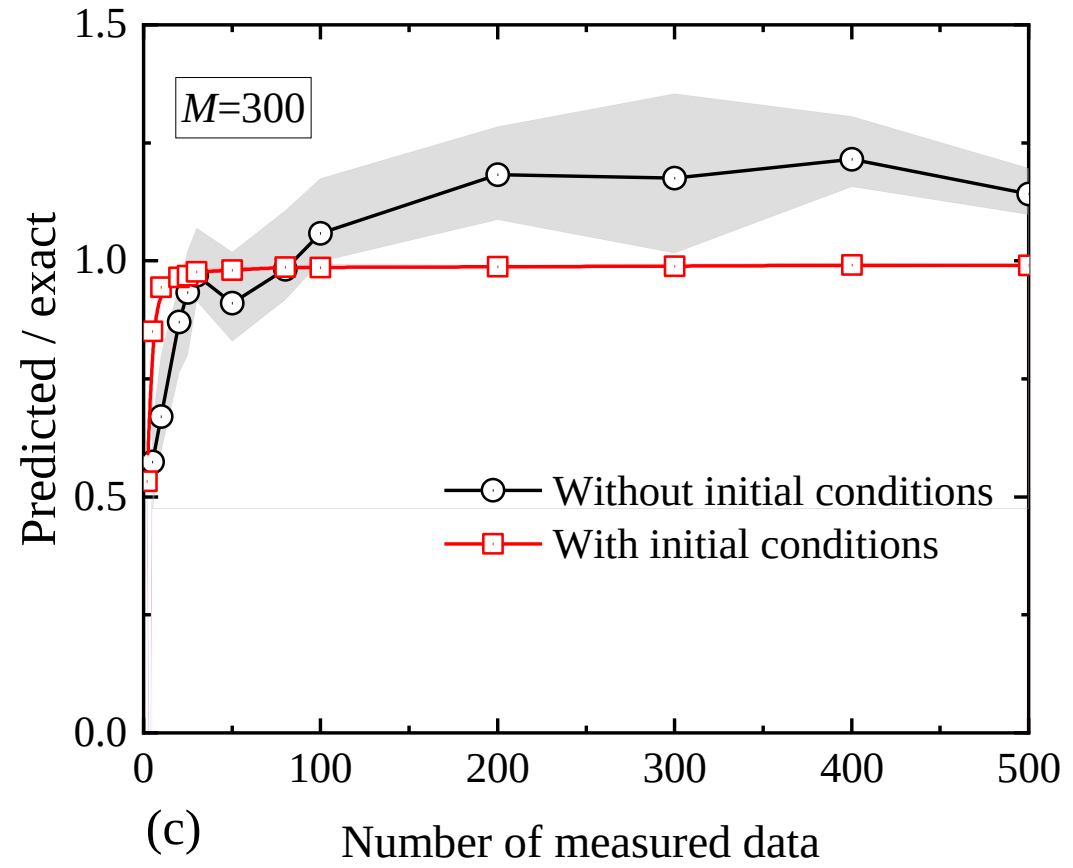

**Figure 14 Back-calculated coefficient of consolidation with and without initial conditions: (a) *M*=100; (a) *M*=200; (a) *M*=300**

After demonstrating the feasibility of PIELM for interpreting CPTu dissipation tests, we present the application of PIELM to real CPTu tests in Lake Alice Clay (Gupta and Davidson 1986) and Boston Blue Clay (Baligh and Levadoux 1980). In both cases, the piezocone probe had a diameter of 38.1 mm and a probe angle of 60°. Laboratory consolidation tests were also performed on the soil samples, and their mean values and variation ranges are plotted in Figure 15, which can provide the benchmark for comparison with CPTu interpretations. To interpret the CPTu dissipation tests using PIELM, the input training points were the same as those listed in Table 3, with *M*=200 and initial conditions removed. Five repeated computations were conducted, and the average values and corresponding error bands are shown as shaded areas in Figure 15. Although considerable scatter exists in both the laboratory and predicted results, it can be seen that the two methods exhibit reasonable agreement with each other.

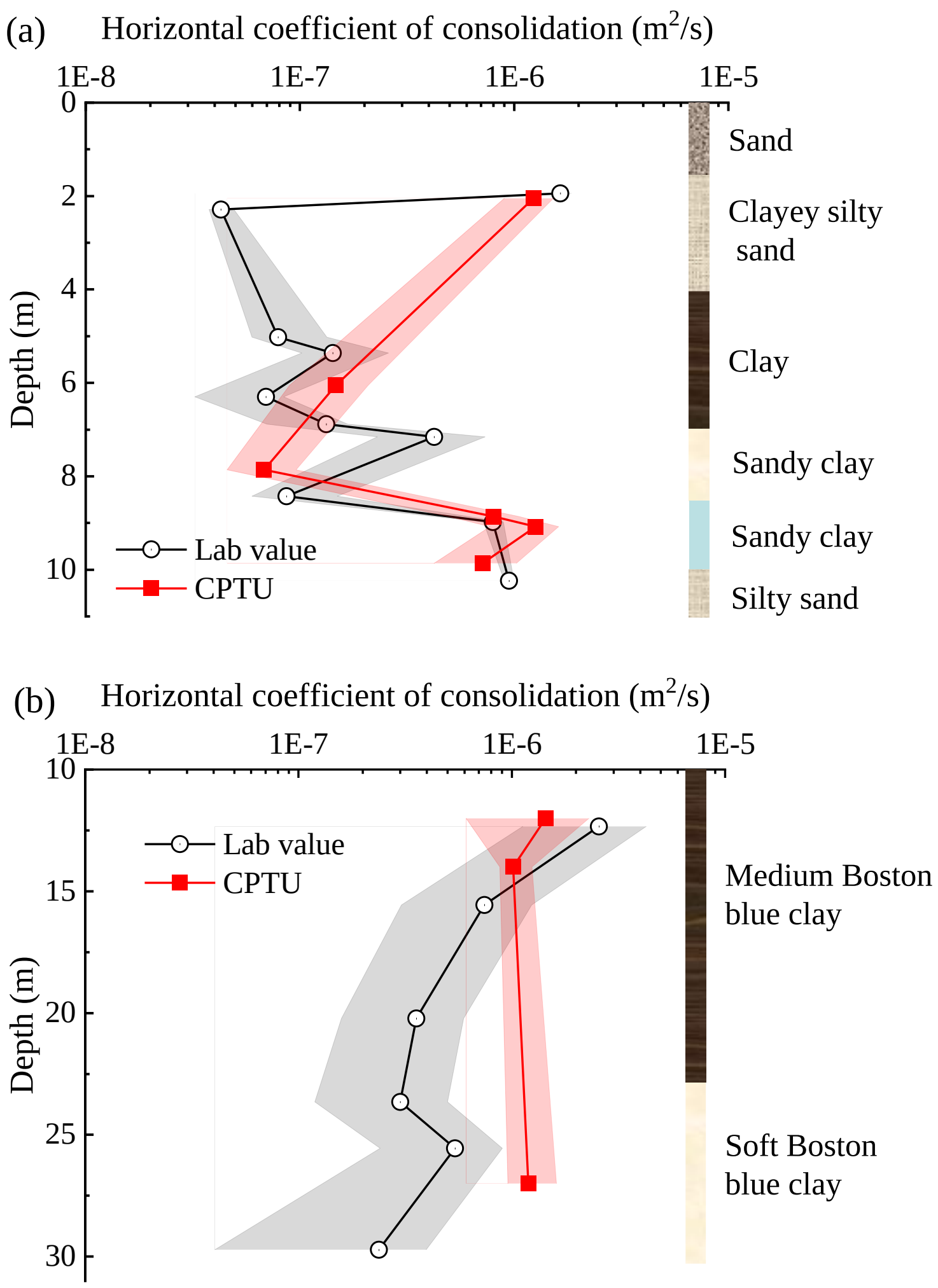

**Figure 15 Comparison of laboratory and predicted coefficients of consolidation:**

**(a) Lake Alice Clay; (b) Boston Blue Clay**

## 6. limitations

While the advantages of PIELM in consolidation analysis and CPTu interpretation are demonstrated, certain limitations remain. The dissipation rate of pore water pressure generally decreases over time as the hydraulic gradient diminishes. To enhance efficiency, TS-PIELM could adopt non-uniform time intervals, such as incorporating self-adaptive schemes. Moreover, the performance of PIELM depends on factors such as the distribution and quantity of training data, as well as the number of hidden-layer neurons. Future work should include rigorous error estimation, probabilistic analysis, and well-defined assessment metrics to enhance the interpretability and reliability of PIELM. Then, the accuracy of the inferred consolidation coefficient from CPTu tests hinges on data quality. Although these coefficients derived from both CPTu and laboratory tests often exhibit scatter, future studies should evaluate the robustness of PIELM against noisy measurements to assess its practical applicability. Last but not least, the excess pore water pressure can also dissipate vertically and the ground conditions can be quite complex in practice, which is not considered in this study. This is primarily because accounting for vertical dissipation would require substantially more measured data to train the network, but such data are typically unavailable in practical CPTu field tests. Hopefully, more advanced CPTu equipment and testing methods may be proposed in the future, making it more "realistic" to interpret the of coefficient of consolidation based on CPTu data.

## 7. Conclusions

This paper investigates the feasibility of physics-informed machine learning (PIML) for solving Terzaghi consolidation equation using a more efficient framework termed PIELM. The deep neural network employed in conventional physics-informed neural network (PINN) frameworks is replaced by a single-hidden-layer extreme learning machine (ELM). Both physical laws and measured data are incorporated into a loss vector, making it convenient for the fusion of data and physics. The input-layer weights of the ELM are assigned randomly, the output-layer weights are computed directly by minimizing this loss vector. In this way training the ELM network can be much more efficient after abandoning the time-consuming gradient-descent method. For forward analysis, a time-stepping strategy is adopted by dividing the consolidation process into a sequence of incremental time steps, resulting in the TS-PIELM framework. When initial and boundary conditions are inconsistent, TS-PIELM helps mitigate the issue of excessively large gradients and thereby improves prediction accuracy. For inverse analysis, this paper presents a preliminary study on applying PIELM to CPTu dissipation tests, from which the coefficient of consolidation of soil can be inferred. Since initial conditions are typically unavailable in such field tests, PIELM is particularly well-suited due to its ability to seamlessly integrate partial physical laws with incomplete measurement data. As long as sufficient high-quality data are available, it shows that PIELM can still accurately estimate the coefficient of consolidation from CPTu dissipation curves, even in the absence of explicit initial conditions.

## Appendix Constrained expressions in case studies

To improve the accuracy of PIML, the network architecture can be modified by applying hard constraints, which ensure that initial/boundary conditions are strictly satisfied (Lu et al. 2021; Schiassi et al. 2021; Lan et al. 2024a; Zhang et al. 2024a). Following Schiassi et al. (2021), the solution of the PDE to be determined can be written in a constrained expression comprising:

(a) a function that analytically satisfies the constraints;

(b) a functional comprising a free function $g(\boldsymbol{x},t)$ that is approximated by the deep neural network in PINN and by the ELM in PIELM.

The constrained expression represents the family of all possible functions that satisfy the constraints, transforming constrained problems into unconstrained problems. For the three case studies in this paper, the initial and boundary conditions do not exactly match each other. Hence, the hard constraints are only applied to the boundary conditions, and the constrained expressions are summarised in this appendix. The detailed method for constructing the expressions can be found in Schiassi et al. (2021) and is not repeated here.

In Case 1, the constrained expression is

$$u(z,t)=g(z,t)-g(0,t)-z(\partial g/\partial z)\big|_{z=H} \tag{30}$$

In Case 2, the constrained expression is

$$u(r,t)=g(r,t)-\frac{r-b}{a-b}g(a,t)+\frac{r-a}{a-b}\left[g(b,t)-p_0\right] \tag{31}$$

Finally, in Case 3, the constrained expression is

$$\begin{aligned} u(r,z,t)=&\ g(r,z,t)-g(a,z,t)-(r-a)(\partial g/\partial r)\big|_{r=b} \\ &-g(r,0,t)+g(a,0,t)+(r-a)(\partial g/\partial r)\big|_{z=0} \\ &-zg_z(r,H,t)+zg_z(a,H,t)+(r-a)z\left(\partial^2 g/\partial r\partial z\right)\big|_{r=b,z=H} \end{aligned} \tag{32}$$

## Data Available Statement

Data will be available from the corresponding author upon reasonable request.

## Conflict of interests

The authors declare that there is no known conflict of interest.

## Declaration of generative AI use

We used large language models (LLMs) to help refine the writing and presentation of this paper (e.g., clarifying explanations, improving readability, and rephrasing drafts). All technical content—including problem formulation, method design, theoretical results, and experiments—was conceived, implemented, and validated by the authors.